\documentstyle[12pt]{article}
\def\ii{\'{\char'20}}

\begin{document}

\begin{titlepage}

\title{
\begin{flushright} {\normalsize UB-ECM-PF-93/11} \end{flushright}
\vspace{2cm}
\large\bf Compactification to non-symmetric homogeneous space in
multidimensional Einstein-Yang-Mills theory}
\author{\large\bf Kubyshin  Yu.A.
\thanks{On sabbatical leave from Nuclear Physics Institute, Moscow State
University, 119899 Moscow, Russia}
\thanks{E-mail address: kubyshin@ebubecm1.bitnet} \\
Departament d'Estructura i Constituents de la Mat\'{e}ria \\
Universitat de Barcelona \\
Av. Diagonal 647, 08028 Barcelona, Spain
      \and
 \large \bf Perez Cadenas J.I. \\
Nuclear Physics Institute, Moscow State University,\\
Moscow 119899, Russia }
\date{30 April 1993}
\maketitle

\begin{abstract}
We study ten-dimensional Ein\-stein - Yang - Mills model with the space of
extra dimensions being a non-sym\-met\-ric ho\-mo\-ge\-ne\-ous space with
the in\-va\-ri\-ant met\-ric parametrized by two scales. Dimensional
reduction of the model is carried out and the scalar potential of the
reduced theory is calculated. In a cosmological setting minima of the
potential in the radiation-dominated period that followed the inflation
are found and their stability is analyzed. The compactifying solution is
shown to be stable for the appropriate values of parameters.

\end{abstract}

\end{titlepage}

\section{Introduction}

Since the seminal papers \cite{KK} by Th. Kaluza and O. Klein the idea of
geometric unification of interactions based on the assumption that spacetime
has more than four dimensions has been attracting considerable attention.
The standard assumption in this approach is that multidimensional spacetime
is modelled by the direct product
\begin{equation}
   E_{(4+d)} = M_{(4)} \times I_{(d)},  \label{eq:factoriz}
\end{equation}
where $ M_{(4)}$ is the macroscopic four-dimensional spacetime and $I_{(d)}$
is a $d$-dimensional
compact Riemannian manifold corresponding to the space of extra dimensions,
which is often referred to as the internal space. The characteristic size
$L$ of $I$ should be small enough so that extra dimensions do not show up at
least up to the energies of the order of $1 TeV$. It is also required that the
geometry of spacetime with such structure satisfies the equations of motion.
If this is true then extra dimensions might curl up to a space of small size
at an early stage of the evolution of the Universe as a result of a dynamical
process called spontaneous compactification.

The main idea of introducing the extra dimensions is that they serve the
purpose of unifying different interactions in four dimensions. In the
original papers \cite{KK} Kaluza and Klein considered pure gravity on
five-dimensional spacetime $M^{4} \times S^{1}$, where $M^{4}$ is the
Minkowski space, which led to the electromagnetism
coupled to Einstein gravity after reduction to four dimensions. In order
to get non-abelian gauge fields in the reduced theory one has to consider
internal spaces $I$ with a non-abelian isometry group $G$ \cite{kerner} -
\cite{coquer}. However, this scheme has some difficulties. One of
them is that there are no spontaneous compactification solutions with the
direct product structure $M_{(D)} = M^{4} \times I$, where space $I$ has
non-abelian isometry group. It should be mentioned that situation will be
different if quantum corrections are taken into account.
Spontaneous compactification solutions in multidimensional gravity in the
framework of Vilkovisky-De Witt effective action formalism were analyzed in
series of papers, see for example \cite{sc-quantum}.
Another problem in the standard Kaluza-Klein approach is that after
adding fermions to the multidimensional action it is impossible to get chiral
fermions after reduction to four dimensions \cite{witten}.

Alternatively one may add Yang-Mills fields to $(4 + d)$-dimensional action
\cite{horv-pal}. The Yang-Mills fields give rise to gauge
fields and Higgs fields with a self-interacting symmetry breaking
potential in four dimensions (see \cite{dimred1} - \cite{KMV} for examples and
\cite{KMRV-review}, \cite{KZ-review} for reviews). In fact multidimensional
Einstein-Yang-Mills theories appear quite naturally as a part of the bosonic
sector in supergravity and superstring models (see \cite{DNP}, \cite{GSW} for
reviews). As it was shown by Cremmer and Sherk and Luciani \cite{cremmer}
compactifying solutions exist in such theories which correspond to a
factorization of spacetime of the type (\ref{eq:factoriz}) with a size $L$
of the order of the Planck length, $L \approx L_{Pl} \approx 10^{-33} cm$.
Spontaneous compactification in a cosmological setting was studied in
series of papers (see \cite{dynsc} - \cite{amendola} for example). In this
approach it occurs as a result of evolution in the early Universe when
the scale factor of three spatial dimensions increases up to its observed
macroscopic value while the scale factor $L(t)$ of the internal space
is static or slowly varying and small. In particular, the process
of compactification in Einstein-Yang-Mills theories with the internal
space $I$ being a symmetric homogeneous space $G/H$ was analyzed in
\cite{KRT}, \cite{BKM}. In those papers the geometry of the space of
extra dimensions was described by $G$-invariant metrics characterized by one
scale.

The important issue is that of the stability of the spontaneous
compactification solutions. It has been shown that some of them are
stable against symmetric \cite{dimred2}, \cite{KRT} and general
\cite{stability} small classical fluctuations. In \cite{BKM} the problem
of stability of compactification in a multidimensional Einstein-Yang-Mills
model in the radiation-dominated period, which followed the inflationary
expansion of three spatial dimensions, was studied. In that analysis
temperature was introduced by allowing the three dimensional components
of the strength tensor of gauge fields to be non-zero.

Considering compactification for the class of invariant metrics on a symmetric
space, as a space of extra dimensions, is an analysis in the spirit of
mini-superspace approach. In general we should consider the same problem
for more general geometries. This will also enable us to understand what
we really learn about the process of compactification from the analysis
for the symmetric internal space. The first step in this
direction is to take into account geometries with the same topology
corresponding to a non-symmetric extension of the internal space, which are
described by invariant metrics with more than one scale. It is
known that all extrema of the action in the class of configurations
corresponding to the symmetric internal space $G/H$ will be extrema of
the action in the class of configurations
corresponding to its non-symmetric extension. The important question is
whether the minimum of the potential in the symmetric case remains stable
against non-symmetric fluctuations. Also it is rather interesting
to find non-symmetric vacua of the model and to understand their role in
the spontaneous compactification. The next step, which is not considered
in the present paper, would be to understand whether
the symmetric case was more preferrable dynamically, namely whether the
system tended to evolve to a symmetric configuration and thus to
restore higher symmetry dynamically. If it is so, this will be
a justification of considering symmetric spaces and invariant metrics
on them as "the first approximation" in the analysis of spontaneous
compactification. We are going to address this issue elsewhere.

We will study these questions in case of an Einstein-Yang-Mills model on
ten-dimensional spacetime $M_{(4)} \times I$, where $I$ is the
non-symmetric homogeneous space $I = SO(5)/SU(2) \times U(1)$. Our results
will be extensions of the results of \cite{KMV} and \cite{BKM}. As it was
mentioned above stability of compactification of extra dimensions to
symmetric homogeneous spaces at the post-inflationary period
(with non-zero temperature taken into account) was studied in \cite{BKM}.
It was assumed there that spacetime had the form
\begin{equation}
       E_{(4+d)} = R^{1} \times S^{3} \times G/H,    \label{eq:robert}
\end{equation}
where $R^{1}$ denotes  a timelike direction, $S^{3}$ represents the
three-dimensional space and $G/H$ is the symmetric space of extra dimensions.
In the present paper we extend this analysis by considering compactification
to a non-symmetric homogeneous space $G/H$. In \cite{KMV} solutions of
spontaneous compactification
corresponding to the spacetime structure $M^{4} \times G/H$ ($M^{4}$ is the
Minkowski spacetime) with non-symmetric internal space $G/H = SO(2k~+~1)
/ SU(k)\times U(1)$ were considered. Here we will consider the internal space
of the same type with $k=2$ (we believe that in the cases when $k > 2$ results
are qualitatively the same) but will look for the compactifying solutions
in a cosmological setting, i.e. of the form (\ref{eq:robert}), with
non-static scale factor $a(t)$ of the three-dimensional space. The relation
between spontaneous compactification solutions with $M_{(4)} = M^{4}$ and
compactifying solutions in the cosmological setting with $M_{(4)} = R^{1}
\times S^{3}$ is discussed in the Appendix. We will find all compactifying
solutions in the model under consideration for non-zero temperature and
analyze their role for the compactification of extra dimensions.

The paper is organized as follows. In Sect. 2 we present some general results
from the theory of G-invariant metrics and G-symmetric gauge fields. This is
the set of configurations in which compactifying solutions will be searched.
We also discuss the  action of a dimensionally reduced
Einstein-Yang-Mills theory. In Sect. 3 we consider dimensional
reduction of the model with non-symmetric internal space $G/H = SO(5)/SU(2)
\times U(1)$ and multidimensional gauge group $K=SU(2+m)$. There
nonvanishing three-dimensional components of the gauge field, which
model the radiation that dominates the energy density of the Universe,
will be taken into account. The effective potential $W$
of scalar fields in four dimensions will be calculated. In Sect. 4
we consider this potential in the case of zero temperature, find its extrema
and analyze their stability. These results are compared with those obtained
in \cite{KMV} and \cite{KRT}. Similar analysis but with non-zero contributions
from radiation is carried out in Sect. 5. It is shown
that there are a stable minimum corresponding to compactifying solution with
the symmetry of the symmetric space, two minima, which correspond to
decompactification of all or part of extra dimensions when the temperature
goes to zero, and two saddle points. Sect. 6 contains our conclusions.

\section{Dimensional reduction of Einstein-Yang-Mills theories}

In this section we discuss dimensional reduction of an Einstein-Yang-Mills
theory on $(4+d)$-dimensional spacetime $E_{(4+d)} = M_{(4)} \times G/H$. For
the sake of generality results for the internal space being an arbitrary
compact homogeneous space are presented here. In the following sections we
will restrict our analysis to a particular model with non-symmetric
coset space $G/H = SO(5) / SU(2) \times U(1)$. In a cosmological
setting the four-dimensional part of multidimensional spacetime is often
assumed to be, in large scales, of the closed Friedman-Robertson-Walker type
$M_{(4)} = R^{1} \times S^{3}$ (see (\ref{eq:robert})), we will specify our
formulas to this case in the following sections.

The large-scale dynamics of the multidimensional Universe is described
by the reduced theory. At the post-inflationary period, which we are
going to consider, the bosonic sector dominates. It is believed
that inclusion of fermions will not lead to any qualitative change in the
analysis of spontaneous compactification. Thus the Einstein-Yang-Mills
theory in multidimensional spacetime seems to be a reasonable model to
study spontaneous compactification. The action of such theory
with gauge group $K$ on multidimensional spacetime (\ref{eq:factoriz})
is given by
\begin{eqnarray}
S      & = & S_{E} + S_{YM}, \label{eq:action-gen} \\
S_{E}  & = & \frac{1}{16 \pi \hat{\kappa}}\int_{E_{(4+d)}} d \hat{x}
           \sqrt{-\hat{g}} ( \hat{R} - \hat{\Lambda}), \label{eq:action-E} \\
S_{YM} & = & \frac{1}{8 \hat{e}^{2}} \int_{E_{(4+d)}} d \hat{x}
           \sqrt{-\hat{g}} Tr \hat{F}_{MN} \hat{F}^{MN}, \; \; \;
           M,N = 0,1,\ldots 3+d, \label{eq:action-YM}
\end{eqnarray}
where $\hat{g} = \det (\hat{g}_{MN})$, $R$ is the scalar curvature,
$\hat{\kappa}$, $\hat{\Lambda}$ and $\hat{e}$ are the gravitational,
cosmological and gauge constants in $(4+d)$ dimensions respectively and
$\hat{F}_{MN}$ is the stress tensor of the gauge field $\hat{A}_{M}$.

To study the cosmological evolution of the model associated with the action
(\ref{eq:action-gen}) - (\ref{eq:action-YM}) we restrict ourselves to
spatially homogeneous and (partially) isotropic configurations. In mathematical
terms this means that the metric is invariant and the gauge field is symmetric
under the action of the group $SO(4) \times G$, where $SO(4)$ is the isometry
group of the three-dimensional space $S^{3} = SO(4)/SO(3)$ and $G$ is the
isometry group of the internal space $G/H$ (cf. (\ref{eq:robert})). This class
of configurations will play an important role in our analysis and we are going
to describe them now. Our exposition of the main results of the theory of
symmetric fields will be based on \cite{salam}, \cite{KMRV-review},
\cite{KZ-review}, \cite{KN}.

Let us consider a spacetime $M^{n} \times S/R$, where $M^{n}$ is the
$n$-dimensional Minkowski space, $S$ is a compact Lie group and $R$ is
its closed subgroup (we
will take $M^{n} = R^{1}$ and $S/R = SO(4)/SO(3) \times G/H = S^{3}
\times G/H$). In the construction of $S$-invariant metrics and $S$-symmetric
gauge fields an important role is played by the so-called Cartan $1$-form.
To define it consider some choice of representatives $\sigma (y) \in S$ in
the cosets $y \in S/R$ such that $[\sigma (y)] = y$. $\sigma (y)$ can be
viewed as a (local) section in the principle bundle $S \rightarrow S/R$.
The Cartan $1$-form $\theta$ on $S/R$ is defined as the pullback of the
canonical left-invariant form on the group $S$:
\begin{equation}
  \theta _{(y)} = \sigma ^{-1} (y) d\sigma (y).  \label{eq:pullback}
\end{equation}

The form $\theta$ takes values in ${\cal S} = Lie(S)$, the Lie algebra of the
group $S$. We assume that the space $S/R$ is reductive, i.e. ${\cal S}$ can
be decomposed into the direct sum of linear subspaces ${\cal R} = Lie(R)$ and
its complement ${\cal M}$ as follows
\begin{eqnarray}
{\cal S} & = & {\cal R} \oplus {\cal M},   \label{eq:decomp1} \\
 & & [{\cal R}, {\cal M}] \subset {\cal M}.  \label{eq:decomp2}
\end{eqnarray}
The subspace ${\cal M}$ can be identified with the space tangent to
$S/R$ at the point $o = [R] \in S/R$. In general it is reducible under
the adjoint action of the algebra ${\cal R}$ and can be further decomposed
into irreducible subspaces:
\begin{eqnarray}
  {\cal M} & = & \sum_{p=1}^{k} {\cal M}_{p},  \label{eq:decomp3} \\
  ad({\cal R}) {\cal M}_{p} & \equiv & [{\cal R}, {\cal M}_{p}] \subset
  {\cal M}_{p}.                    \nonumber
\end{eqnarray}
We will assume for simplicity the representations of ${\cal  R}$ in
${\cal M}_{p}$ and ${\cal M}_{p'}$ $(p \neq p')$ to be inequivalent.
In accordance with (\ref{eq:decomp1}), (\ref{eq:decomp2}) the Cartan form
(\ref{eq:pullback}) can be decomposed as
\begin{eqnarray}
   & \theta & = \theta _{{\cal R}} + \sum_{p=1}^{k} \theta _{{\cal M}_{p}},
   \label{eq:theta1} \\
  \theta _{{\cal R}}  =  \sum_{\bar{a} = 1}^{\dim R}
  (& \theta _{{\cal R}})^{\bar{a}} &
   T_{\bar{a}} , \; \; \; \;
   \theta _{{\cal M}_{p}} = \sum_{i=1}^{d_{p}} \theta ^{p,i} T_{p,i},
   \label{eq:theta2}
\end{eqnarray}
where 1-forms $\theta _{{\cal R}}$ and $\theta _{{\cal M}_{p}}$ take values in
${\cal R}$ and ${\cal M}_{p}$ respectively, $\{ T_{\bar{a}}, \bar{a} = 1,
 \ldots , \dim R \}$ is a basis in the Lie algebra ${\cal R}$, $d_{p} = \dim
{\cal M}_{p}$ and $\{ T_{p,i}, i = 1, \ldots , d_{p} \}$ is a basis in the
subspace ${\cal M}_{p}$. We choose $\{ T_{p,i} \}$ to be orthonormal with
respect to an $ad R$-invariant scalar product in ${\cal M}$.

The 1-forms $\theta ^{p,i}$ form a local moving co-frame on $S/R$. In this
co-frame the components of an $S$-invariant metric on $S/R$ are independent of
the local coordinates $y^{m}$ $(m=1, \ldots , d = \dim {\cal M})$ and its
most general form for the appropriate choice of the basis in ${\cal M}$
reads
\begin{equation}
 \gamma _{(y)} = \sum_{p=1}^{k} L_{p}^{2} \sum_{i=1}^{d_{p}}
   \theta _{(y)}^{p,i} \theta _{(y)}^{p,i}.   \label{eq:dmetric}
\end{equation}

Then the $S$-invariant metric on $M^{n} \times S/R$ at point $\hat{x}$
is given by
\begin{equation}
 \hat{g} _{MN} (\hat{x}) = \left( \begin{array}{cc}
                                   g_{\mu \nu} (x) & 0 \\
                                   0 & \gamma _{mn}(x,y) \end{array}
                            \right) ,   \label{eq:gmetric}
\end{equation}
where $\hat{x} = (x^{\mu},y^{m})$, $x \in M^{n}$, $y \in S/R$,
$g_{\mu \nu}(x)$ is an arbitrary metric on $M^{n}$. The metric
components $\gamma _{mn}$ are determined by (\ref{eq:dmetric}) with $L_{p}$
depending on $x$, $L_{p} = L_{p}(x)$, which have the dimension of length
and characterize the "sizes" of the internal space at point $x$.

Let us now discuss $S$-symmetric gauge fields on $S/R$. Here $S$-symmetry means
that the fields are invariant under $S$-transformations up to a gauge
transformation. Thus, both the group $S$ of spatial symmetry and the gauge
group $K$ are involved in this definition. Indeed, an $S$-symmetric gauge
field on $M^{n} \times S/R$ is characterized by a homomorphism $\lambda$ of
the isotropy group $R \subset S$ to the gauge group $K$: $\lambda : R
\rightarrow K$, and its components $\hat{A}_{M}(\hat{x})$ can be read from
the 1-form $\hat{A}=\hat{A}_{M}(\hat{x}) d\hat{x}^{M}$ given by
\begin{equation}
\hat{A} = A_{\mu} dx^{\mu} + \lambda(\theta_{{\cal R}}) +
          \phi _{x}(\theta _{{\cal M}}).    \label{eq:symfield}
\end{equation}
This formula is the result of Wang's theorem \cite{KN}. Here $A_{\mu}(x)$
is a gauge field on $M^{n}$ with the gauge group $C = \{ c \in K |
\lambda (r) c \lambda (r) ^{-1} = c \; \; \mbox{for all} \; \; r \in R \}
\subset K$, the centralizer of $\lambda (R)$ in $K$. $\phi _{x}$ in
(\ref{eq:symfield}) is a mapping from ${\cal M}$ to ${\cal K} = Lie(K)$,
$\phi _{x} : {\cal M} \rightarrow {\cal K}$, which is an intertwining
operator. Its structure is determined by the irreducible representations of
${\cal R}$ in ${\cal M}$ and ${\cal K}$, and it can be constructed as follows.
Let us, in addition to the decomposition (\ref{eq:decomp3}), also decompose
${\cal K}$ into irreducible representations with respect to $ad \lambda (R)$.
Suppose that we have
\begin{equation}
 {\cal K} = \lambda ({\cal R}) \oplus {\cal C} \oplus \sum _{r=1}^{n}
 \sum _{i=1}^{n_{r}} V_{i}^{(r)} \oplus \Xi, \label{eq:kdecomp}
\end{equation}
where ${\cal C} = Lie(C)$,  $V_{i}^{(r)}  \; \; (i = 1, \ldots , n_{r})$ for
a given $r$ $(r = 1, \ldots , n)$ are linear subspaces carrying the
irreducible representation of ${\cal R}$ equivalent to that in ${\cal M}_{r}$
and the subspace $\Xi$ carries other irreducible representations none of which
is equivalent to those in subspaces ${\cal M}_{p}$ in (\ref{eq:decomp3}). It is
clear that $n \leq k$. Then the mapping $\phi _{x}$ is equal to
\begin{eqnarray}
  \phi_{x} & = & \sum _{p=1}^{n} \phi _{x}^{(p)}, \label{eq:phi-mapping} \\
  \phi _{x}^{(p)} (u) & = & \sum _{i=1}^{n_{p}} f_{i}^{(p)}(x) \phi _{i}^
   {(p)} (u_{p}),    \nonumber
\end{eqnarray}
where $u \in {\cal M}$, $u_{p}$ is the component of the vector $u$ in
${\cal M}_{p}$ and $\phi _{i}^{(p)}$ is an elementary intertwining operator
establishing the homomorphism between subspaces ${\cal M}_{p} \subset
{\cal M} \subset {\cal S}$ and $V_{i}^{(p)} \subset {\cal K}$ carrying
equivalent irreducible representations of ${\cal R}$. We will see shortly that
the functions $f_{i}^{(p)} (x)$ are scalar fields in the reduced theory.
They form multiplets $f ^{(p)} = \{ f_{1}^{(p)}, \ldots , f_{n_{p}}^{(p)} \}$
(in general reducible) with respect to the action of the gauge
group $C$ of the reduced theory.

Let us return now to the problem of dimensional reduction of the
multidimensional Einstein-Yang-Mills action
(\ref{eq:action-gen})-(\ref{eq:action-YM}) on $M_{(4)} \times G/H$
without specifying the type of four-dimensional spacetime for the time being.
Using eqns. (\ref{eq:dmetric}) - (\ref{eq:symfield}) and
(\ref{eq:phi-mapping}) we carry out the dimensional reduction of the
gauge part of the action (\ref{eq:action-YM}) and get the following result
\cite{horv-pal}-\cite{dimred1}, \cite{KMRV-review}-\cite{KZ-review}
\begin{eqnarray}
  S_{YM} & = & \frac{1}{8 e^{2}} \int _{M_{(4)}} d^{4}x \sqrt{-g_{(4)}}
  (\frac{L_{1}(x)}{L_{10}})^{d_{1}} \ldots (\frac{L_{k}(x)}{L_{k0}})^{d_{k}}
  \nonumber  \\
  & \times &  [ Tr ( F_{\mu \nu } F^{\mu \nu} ) - \sum_{p=1}^{n} \frac{d_{p}}
  {L_{p}^{2}} Tr | D_{\mu} f^{(p)} (x) |^{2} - V(f;L_{p}) ],
  \label{eq:reduced-YM}
\end{eqnarray}
where $F_{\mu \nu}$ and $D_{\mu} f^{(p)}$ are respectively the stress tensor
of the gauge field $A_{\mu}(x)$ and the gauge covariant derivative of the
scalar multiplet $f^{(p)}(x)$ (see (\ref{eq:symfield})) and
$g_{(4)} = \det(g_{\mu \nu})$. The gauge coupling
constant of the reduced theory is given by
\begin{equation}
  e ^{2} = \hat{e}^{2} / v_{0}(G/H),    \label{eq:coupling}
\end{equation}
where $v_{0}(G/H)$ is the volume of the internal space with characteristic
sizes being equal to $L_{10}, \ldots , L_{k0}$, parameters to be
specified later. $V(f)$ is a quartic self-interaction potential of the
scalar fields equal to
\begin{eqnarray}
V(f;L_{p}) & = & \sum_{p,q = 1}^{n} \frac{1}{L_{p}^{2}} \frac{1}{L_{q}^{2}}
   \sum _{i=1}^{d_{p}} \sum _{j=1}^{d_{q}} Tr (\hat{F}_{(p,i)(q,j)}
   \hat{F}_{(p,i)(q,j)} ),  \label{eq:potential-gen}  \\
 \hat{F}_{(p,i)(q,j)} & = & [ \phi _{x} (T_{p,i}), \phi _{x} (T_{q,j}) ] -
     \lambda ([T_{p,i}, T_{q,j}]_{{\cal H}}) -
     \phi _{x} ([T_{p,i},T_{q,j}]_{{\cal M}}),  \nonumber
\end{eqnarray}
where similar to previous notations ${\cal H} = Lie(H)$ and ${\cal G} = Lie(G)=
{\cal H}  \oplus {\cal M} $. Thus, the theory (\ref{eq:reduced-YM}) -
(\ref{eq:potential-gen}), obtained by dimensional reduction of
(\ref{eq:action-YM}) is the theory of gauge fields with the gauge group
$C \subset K$ coupled to the scalar fields $f^{(p)}(x)$. Explicit
calculations of the symmetric gauge 1-form (\ref{eq:symfield}) and the
intertwining mapping (\ref{eq:phi-mapping}) as well as the scalar potential
(\ref{eq:potential-gen}) can be found in \cite{horv-pal} -
\cite{KZ-review}, \cite{BKM}, \cite{BMPV}.

Let us now discuss dimensional reduction of the Einstein part $S_{E}$ of
the multidimensional action (\ref{eq:action-E}). Using (\ref{eq:dmetric}),
(\ref{eq:gmetric}) and performing rather straightforward though lengthy
calculations one gets
\begin{eqnarray}
 S_{E} & = & \int_{M_{(4)}} d^{4}x \sqrt{-g_{(4)}} (\frac{L_{1}(x)}{L_{10}})^
   {d_{1}} \ldots (\frac{L_{k}(x)}{L_{k0}})^{d_{k}} \nonumber \\
       & \times & \{ \frac{1}{16 \pi \kappa} [ \tilde{R}^{(4)} +
    \sum _{p=1}^{k} \frac{d_{p}}{L_{p}^{2}} ( R_{p} - g^{\mu \nu}
    \frac{\partial_{\mu} L_{p}(x) \partial_{\nu} L_{p}(x)}{L_{p}^{2}(x)} )
     \nonumber \\
       & + & \sum_{p,q = 1}^{k} g^{\mu \nu} d_{p} d_{q}
       \frac{\partial_{\mu} L_{p}(x) \partial_{\nu} L_{q}(x)}{L_{p}(x)
       L_{q}(x)} ]-\hat{\Lambda} v_{0}(G/H)\}, \label{eq:reduced-E}
\end{eqnarray}
where
\begin{equation}
  \kappa  = \hat{\kappa} / v_{0}(G/H),   \label{eq:kappa}
\end{equation}
$\tilde{R}^{(4)}$ is the scalar curvature in four dimensions calculated with
respect to the metric $g_{\mu \nu}(x)$. The coefficients $R_{p}$ of the
curvature on $G/H$ were calculated using formulas from ref. \cite{jadczyk}
(see also \cite{KMV}, \cite{mourao} and the book \cite{KMRV-review}):
\begin{eqnarray}
 R_{p} & = & \frac{1}{2d_{p}} \sum_{q,t=1}^{k} (\frac{1}{2} \frac{L_{p}^{4}}
 {L_{q}^{2} L_{t}^{2}} - \frac{L_{t}^{2}}{L_{q}^{2}} ) {\cal D}_{qt}^{p} +
  \frac{1}{2} C_{2}^{G} (Ad), \label{eq:Rp}  \\
  {\cal D}_{qt}^{p} & = & \sum_{i=1}^{d_{p}} \sum_{j=1}^{d_{t}} \sum_{m=1}^
  {d_{p}} [C_{(q,i)(t,j)}^{(p,m)} ]^{2}. \nonumber
\end{eqnarray}
Here $C_{(q,i)(t,j)}^{(p,m)}$ are the structure constants of the group $G$
in the basis $\{ T_{\bar{a}}, T_{p,i} \}$ introduced in
(\ref{eq:theta1}), (\ref{eq:theta2}) and $C_{2}^{G} (Ad)$ is the eigenvalue
of the quadratic Casimir operator of $G$ in the adjoint representation.

For the purpose, which will be seen shortly, we perform the following conformal
transformation of the four-dimensional metric:
\begin{equation}
  g_{\mu \nu}(x) = (\frac{L_{1}(x)}{L_{10}})^
   {d_{1}} \ldots (\frac{L_{k}(x)}{L_{k0}})^{d_{k}} \eta _{\mu \nu}(x).
   \label{eq:conformal}
\end{equation}
Then the action (\ref{eq:reduced-E}) takes the form
\begin{eqnarray}
 S_{E} & = & \int_{M_{(4)}} d^{4}x \sqrt{-\eta} \{ \frac{1}{16 \pi \kappa}
  [ R^{(4)} - \frac{1}{2} \sum_{p,q=1}^{k} \eta ^{\mu \nu}  Q_{pq}
  \frac{\partial_{\mu} L_{p} \partial_{\nu} L_{q}}{L_{p}L_{q}}
  \label{eq:reduced2-E} \\
  & + &  (\frac{L_{10}}{L_{1}})^{d_{1}} \ldots (\frac{L_{k0}}{L_{k}})^{d_{k}}
  \sum_{p=1}^{k} \frac{d_{p}}{L_{p}^{2}} R_{p} ] -\hat{\Lambda} v_{0}(G/H)
  \prod _{p=1}^{k} (\frac{L_{p0}}{L_{p}(x)})^{d_{p}} \},   \nonumber
\end{eqnarray}
where $\eta = \det(\eta _{\mu \nu})$, $R^{(4)}$ is the scalar curvature
in four dimensions calculated with respect to the metric $\eta _{\mu \nu}$
and
\begin{equation}
Q _{pq} = 2 d_{p} \delta _{pq} + d_{p} d_{q}.   \label{eq:Q-matrix}
\end{equation}
We see that after conformal rotation (\ref{eq:conformal}) and changing
variables
\begin{equation}
  L_{p}(x) = L_{p0} e^{\psi _{p} (x)}   \label{eq:psi-def}
\end{equation}
the gravitational part of the multidimensional action (\ref{eq:action-gen})
after dimensional reduction can be interpreted as Einstein gravity
coupled to scalar fields with exponential potential.

Summing up (\ref{eq:reduced-YM}) and (\ref{eq:reduced2-E}) and taking eqns.
(\ref{eq:conformal}) and (\ref{eq:psi-def}) into account we get the
formula for the action of the theory on $M_{(4)}$ obtained by dimensional
reduction of a Einstein-Yang-Mills model on $(4+d)$-dimensional spacetime
$E_{(4+d)} = M_{(4)} \times G/H$:
\begin{eqnarray}
S & = & \int_{M_{(4)}} d^{4}x \sqrt{-\eta} \{ \frac{1}{16 \pi \kappa}
 [R^{(4)} - \frac{1}{2} \sum_{p,q=1}^{k} \eta^{\mu \nu} Q_{pq} \partial_{\mu}
 \psi _{p} \partial_{\nu} \psi_{q} ]      \nonumber \\
 & + &  \frac{1}{8 e^{2}} [ e^{\sum_{p=1}^{k} d_{p} \psi_{p}}
     Tr F_{\mu \nu} F^{\mu \nu} - \sum_{q=1}^{n} d_{q} L_{q0}^{2}
     e^{-2 \psi_{q}} Tr D_{\mu} f^{(q)} (D^{\mu} f^{(q)})^{*} ]
   \nonumber \\
   &  & - W(\psi_{p}, f^{(q)}) \},      \label{eq:DR-action}
\end{eqnarray}
where the potential of scalar fields $\psi _{p}(x)$ $(p=1, \ldots , k)$ and
$f^{(q)}(x)$ $(q=1, \ldots , n \leq k)$ is given by
\begin{equation}
W(\psi_{p}, f^{(q)}) = e^{-\sum_{p=1}^{k} d_{p}\psi_{p}(x)}
 [-\sum_{p=1}^{k} \frac{d_{p}}{16 \pi \kappa} \frac{R_{p}}{L_{p0}^{2}}
 e^{-2 \psi_{p}} + \frac{\hat{\Lambda}}{16 \pi \kappa} + \frac{V(f;L_{p})}
 {8 e^{2}} ].      \label{eq:potential-W}
\end{equation}
Vacua with compactified extra dimensions are static solutions
of the classical equations of motion with finite $\psi_{p}$.

If we neglect contributions from the four-dimensional components of the
gauge field then the compactifying solutions with constant $f^{(q)}$ and
$\psi_{p}$ are given by the extrema of the potential $W$. When
four-dimensional spacetime is assumed to be the Minkowski spacetime then such
solutions are usually referred to as spontaneous compactification solutions.
When $M_{(4)}$ is the closed Friedman-Robertson-Walker universe and
contributions from nonvanishing components of $A_{\mu}$ are taken into
account, compactifying solutions are determined by the potential
$\tilde{W}$ which is the sum of $W$ and two last terms in the square
brackets in (\ref{eq:DR-action}) which model contributions from radiation.
The term proportional to $Tr(F_{\mu \nu} F^{\mu \nu})$ changes the
asymptotic behaviour of the effective potential of the fields
$\psi_{p}$ for large $|\psi| = \sqrt{\psi_{1}^{2} + \ldots + \psi_{k}^{2}}$
and is important for the stability of the vacua. In the next
section contribution of this term to the effective potential will be
calculated explicitly for a certain model.

Eqns. (\ref{eq:DR-action}) and (\ref{eq:potential-W}) generalize
the results obtained in \cite{KRT}, \cite{BKM} where the reduction of the
Einstein-Yang-Mills model with $G/H$ being a symmetric space was considered.
For a symmetric space a $G$-invariant metric (\ref{eq:dmetric}) on $G/H$
is characterized by one scale $L(x)$, so that $k=1$ in (\ref{eq:DR-action}),
(\ref{eq:potential-W}).

The kinetic term for the scalar fields $\psi_{p}(x)$ in (\ref{eq:DR-action})
is not diagonal and physical fields are combinations of $\psi_{p}$
which diagonalize it. This is important if issues of dynamics are
addresses, however diagonalization is not necessary for our purpose.
Notice also that functions $\psi_{p}(x)$ correspond to scalar fields
arising from the metric in the standard Kaluza-Klein theories,
when pure gravity in multidimensional spacetime is considered.

\section{Dimensional reduction in the Einstein-Yang-Mills model with
$G/H = SO(5)/SU(2) \times U(1)$.}

In this section we will consider the Einstein-Yang-Mills theory on
spacetime $E_{(10)} = M_{(4)} \times G/H$, where the space of extra
dimensions is the non-symmetric homogeneous space $G/H =SO(5)/SU(2)
\times U(1)$ with $G$-invariant metric characterized by two scales $L_{1}$
and $L_{2}$. In this case $\dim G/H = 6$ and $\mbox{rank} G =
\mbox{rank} H$ which makes the model relevant for the problem of
compactification in superstrings. We will take the gauge group of the
multidimensional Yang-Mills field to be $K = SU(2+m)$ $(m \geq 3)$.
As it was mentioned in the Introduction dimensional reduction of the Yang-Mills
sector of this model and the scalar potential $V(f)$ (\ref{eq:potential-gen})
for $m=3$ were studied in ref. \cite{FKSZ}. Spontaneous compactification
solutions for this model for arbitrary $m \geq 3$ were found in \cite{KMV}.

We will begin by recovering the results of \cite{FKSZ}, \cite{KMV} concerning
the dimensional reduction in this model with arbitrary four-dimensional
spacetime $M_{(4)}$. Our next step will be to consider the
model in a cosmological setting when the four-dimensional part of the
Universe is of the closed Friedman-Robertson-Walker type $M_{(4)} = R
\times S^{3}$.

The decomposition (\ref{eq:decomp3}) of the linear space ${\cal M}$ after
complexification takes the form
\begin{eqnarray}
  {\cal M} & = & {\cal M}_{1} \oplus {\cal M}_{2},  \; \; \;
  \dim {\cal M}_{1} = d_{1} = 4,   \; \; \; \dim {\cal M}_{2} = d_{2} = 2,
    \nonumber \\
  {\cal M}_{1} & = & {\cal U}_{1} [\underline{2} (1)] + \bar{{\cal U}}_{1}
   [\underline{2} ^{*} (-1)],  \label{eq:decomp-M}  \\
  {\cal M} _{2} & = & {\cal U}_{2} [\underline{1} (2)] + \bar{{\cal U}}_{2}
   [\underline{1} (-2)],  \nonumber
\end{eqnarray}
where the bar over ${\cal U}_{p}$ means complex conjugation and in square
brackets we indicate the types of the irreducuble representations carried by
the subspaces ${\cal U}_{p}$ and $\bar{{\cal U}}_{p}$, $(p=1,2)$ with
respect to $H=SU(2) \times U(1)$.
According to the theory of invariant metrics, sketched in Sect. 2, the
most general $SO(5)$-invariant metric is given by
(\ref{eq:dmetric})-(\ref{eq:gmetric}) and characterized by two scales. We
denote the scale corresponding to ${\cal M}_{1}$ by $L_{1}$ and the one
corresponding to ${\cal M}_{2}$ by $L_{2}$. The homogeneous space $G/H =
SO(5)/SU(2) \times U(1)$ with such metric is known as the squashed $CP^{3}$
manifold, its properties were discussed in detail in ref. \cite{ziller}.
Often this space is represented in another equivalent form as $Sp(2)/SU(2)
\times U(1)$. When
\begin{equation}
 L_{2}^{2} = 2 L_{1}^{2}     \label{eq:L-relation}
\end{equation}
the metric becomes the Einstein metric (the standard Fubini-Study metric)
with $SU(4)$ symmetry of the symmetric space $SU(4)/SU(3) \times U(1)$
corresponding to non-squashed $CP^{3}$. The second Einstein metric
with $L_{1}^{2} = L_{2}^{2}$ (the so called normal metric) does not
correspond to any enlargement of the initial $SO(5)$-symmetry.

Let us consider now the $SO(5)$-symmetric gauge field configurations. We take
the regular embedding $\lambda : H=SU(2) \times U(1) \rightarrow K=SU(m+2)$
defined by the following branching rule for the fundamental representation
$\underline{m+2}$ of $K$:
\[ \underline{m+2} \downarrow _{\lambda (H)} = \underline{2} (1) +
    m \; \; \underline{1} (0)  \]
(we indicate the eigenvalue with respect to the $U(1)$ subgroup of
$\lambda(H)$ in brackets). Then the gauge group of the reduced theory
is $C = SU(m) \times U(1)$ and only the irreducible representations
$[\underline{2} (1)]$ and $[\underline{2}^{*} (-1)]$ in (\ref{eq:decomp-M})
are intertwined by the operator $\phi _{x}$ with the representations in the
decomposition analogous to (\ref{eq:kdecomp})  \cite{KMV}. Thus
$n=1$ in eqns. (\ref{eq:kdecomp}), (\ref{eq:phi-mapping}). The
multiplet $ f(x) \equiv f^{(1)}(x) = \{ f_{1}^{(1)}(x), \ldots ,
f_{m}^{(1)}(x) \}$ of the scalar fields, given by the extra dimensional
components of the gauge field, transforms as a vector in the fundamental
representation of $SU(m) \subset C$.

The potential $V(f;L_{p})$ of the scalar fields obtained by dimensional
reduction of the gauge sector of the multidimensional theory is given by
\cite{KMV}
\begin{equation}
V(f; L_{1},L_{2}) = \frac{12}{L_{1}^{4}} \left[ \left( |f|^{4} -
      \frac{2 - \beta }{3} \right)^{2} +
     \frac{2m + 1}{9(m+2)} - \frac{4}{9} \beta + \frac{2 (m-1)}{9(m+2)}
     \beta ^{2} \right] ,      \label{eq:potential-model}
\end{equation}
where $\beta = L_{1}^{2} / L_{2}^{2}$.
As it was pointed out in \cite{KMV} this expression displays the relation
between the existence of the spontaneous symmetry breaking phenomenon
in the reduced theory and the value of $\beta$
characterizing the geometry of the space of extra dimensions. If
$\beta < \beta_{cr} = 2$ the potential (\ref{eq:potential-model}) is of
the Higgs type and the Higgs vacuum is given by $|\tilde{f}|^{2} =
(2-\beta)/3 $. For $\beta \geq \beta_{cr}$ the potential has only trivial
minimum $|\tilde{f}|= 0$. This is summarized in the following formula
for the value of the scalar field minimuzing the potential $V(f;L_{1},L_{2})$:
\begin{equation} |\tilde{f}|^{2} = \left\{ \begin{array}{ll}
                                      0 & \mbox{if $\beta \geq 2$} \\
                                      \frac{2 - \beta}{3}&
				      \mbox{if $\beta < 2$}.
                                      \end{array}
                                      \right.    \label{eq:f-minimum}
\end{equation}

The quantities $R_{p}$ $(p=1,2)$ in eq. (\ref{eq:reduced-E}),
(\ref{eq:potential-W}) are easily calculated using the formulas (\ref{eq:Rp})
and are equal to
\begin{equation}
 R_{1}(\beta) = 1 - \frac{1}{2\beta}, \; \; \; \;
 R_{2}(\beta) = \frac{2}{\beta ^{2}} - 3.    \label{eq:Rp-model}
\end{equation}
The action of the Einstein-Yang-Mills theory under consideration, after
reduction to four dimensions, takes the form (\ref{eq:DR-action}),
(\ref{eq:potential-W}) with the potential $V(f;\psi_{1}, \psi_{2})$ given
by the formula (\ref{eq:potential-model}) and describes the interaction of
four-dimensional Einstein gravity (metric $\eta _{\mu \nu}(x))$,
non-abelian gauge field $A_{\mu}(x)$ and the scalar fields $f_{i}(x)$,
$\psi_{1}(x)$ and $\psi_{2}(x)$.

In the cosmological setting the four-dimensional spacetime is assumed to have
at large scales the form $M_{(4)} = R^{1} \times S^{3} = R^{1} \times
SO(4)/SO(3)$. To study the cosmological model associated with the theory
considered here we restrict ourselves to spatially homogeneous and (partially)
isotropic field configurations. This means that we consider gauge field
configurations $A_{\mu}$ which are $SO(4)$-symmetric, a metric
$\eta_{\mu \nu}$ which is $SO(4)$-invariant and scalar fields $f_{i}$,
$\psi_{p}$ which depend on time $t$ only. Then all physical quantities are
spatially homogeneous and isotropic. To describe such configurations and to
carry out further reduction of the action we will use the results from the
theory of symmetric fields, outlined in the previuos section, applying
them now to the coset space $S^{3} = SO(4)/SO(3)$. Simple analysis
shows that the most general form of an $SO(4)$-invariant metric reads
\begin{equation}
  \eta = -N^{2}(t) dt^{2} + a^{2}(t) \sum_{a=1}^{3} \theta ^{a}(\vec{x})
   \theta^{a}(\vec{x}),   \label{eq:4-metric}
\end{equation}
where the one-forms $\theta^{a}$ are the components
of the Cartan form on $S^{3}$. They form a moving coframe on $S^{3}$.
The scale $a(t)$ of the three-dimensional space and the lapse function
$N(t)$ are arbitrary nonvanishing functions of time (cf. (\ref{eq:dmetric})).

In our approach contributions to the temperature coming from radiation
are modelled by non-vanishing three-dimensional components of the gauge
field. For the sake of simlicity we will calculate these contributions
for $m=3$ only, the generalization to $m > 3$ being straightforward
(recall that $m$ is the integer determining the rank of the multidimensional
gauge group $K$ and of the rank of the gauge group $C$ of the reduced theory).
To fix $SO(4)$-symmetric gauge fields on $S^{3}$ we need to choose a
homomorphism of the isotropy group $SO(3)$ of the sphere into the gauge
group $C=SU(3) \times U(1)$ of the four-dimensional reduced theory (see
(\ref{eq:symfield}) and the paragraph before that formula). We denote it
by $\tau$ and define it to be an embedding determined
by the following requirement: the fundamental representation of $SU(3)
\subset C$ restricted to the subgroup $\tau(SO(3)) \subset SU(3)$
is irreducible,
\[ \underline{3} \downarrow _{\tau (SO(3))} = \underline{3}. \]
It can be shown then that the symmetric gauge field is characterised by the
function $B(t) = A_{0}(t)$ and one scalar field $h(t)$ generated by
three-dimensional components of the gauge field. The term quadratic in
$F_{\mu \nu}$ in the action (\ref{eq:DR-action}) is equal to
\begin{equation}
 - Tr (F_{\mu \nu} F^{\mu \nu}) =
\frac{6}{N^{2}(t)} \frac{\dot{h}^{2}
(t)}{a^{2}(t)} - \frac{6}{a^{4}(t)} (1-h^{2}(t))^{2},   \label{eq:F2-term}
\end{equation}
whereas the covariant derivative of the scalar field takes the form
\begin{equation}
 - Tr |D_{\mu} f |^{2} = \frac{1}{N^{2}(t)} |D_{0}  f|^{2} -
   \frac{(1 + h^{2}(t))}{a^{2}(t)} |f(t)|^{2},  \label{eq:covder}
\end{equation}
where
\[   D_{0} f = \dot{f} + Bf.   \]
Substituting these expressions to eq. (\ref{eq:DR-action}) we obtain
the following effective action in four dimensions
\begin{eqnarray}
S & = & 16 \pi^{2} \int dt N a^{3} \{ - \frac{3}{8 \pi \kappa} \frac{1}{a^{2}}
 (\frac{\dot{a}}{N})^{2} + \frac{3}{32 \pi \kappa}
 \frac{1}{a^{2}} + \frac{1}{2} \sum_{p,q=1}^{2} Q_{pq}
 (\frac{\dot{\psi}_{p}}{N})^{2}    \nonumber   \\
  & + & L_{10}^{2}  e^{-2 \psi_{1}} \frac{2 |D_{0}f(t)|^{2}}{N^{2}(t)}
   + \frac{3}{4 e^{2}} \frac{e^{4 \psi_{1} + 2 \psi_{2}}}{a^{2}(t)}
   \frac{(\dot{h}(t))^{2}}{N^{2}(t)}
   - \tilde{W}(\psi_{p},f,h) \},  \label{eq:DR-model}
\end{eqnarray}
where the scalar potential is the sum of the term $W(\psi_{p},f)$
(\ref{eq:potential-W}) and the contributions  from the three-dimensional
components of the gauge field given by (\ref{eq:F2-term}), (\ref{eq:covder}):
\begin{equation}
\tilde{W}(\psi_{p},f,h) = W(\psi_{p},f) + \frac{3}{a^{4}(t)}
 e^{4 \psi_{1} + 2 \psi_{2}} \frac{(1-h^{2}(t))^{2}}{4 e^{2}} +
 \frac{2}{a^{2}(t)}L_{10}^{2} e^{-2 \psi_{1}} (1+h^{2}(t))^{2} |f(t)|^{2}.
 \label{eq:W-model}
\end{equation}
Notice that the effective action does not depend on time derivatives of
$N$ and $B$. This means that
these variables play the role of Lagrange multipliers associated with local
symmetries of the effective Lagrangian.

The evolution of the multidimensional Universe described by the model is
determined by the equations of motion for the functions $a(t)$, $\psi_{1}(t)$,
$\psi_{2}(t)$, $f_{i}(t)$ and $h(t)$ which can be obtained from
(\ref{eq:DR-model}), (\ref{eq:W-model}). We will analyse
solutions of these equations elsewhere. In the next sections of the present
paper the problem of stability of compactifying vacua will be considered
and some qualitative conclusions about compactification of extra dimensions
will be discussed.

\section{Stability of compactification without temperature corrections}

Let us first find the vacua corresponding to compactification of extra
dimensions and analyse their stability (in fact, we will see that there
is only one such vacuum) in the case when all contributions to the
temperature, which are modelled by non-vanishing three-dimensional
components of the gauge field, are neglected. This amounts to dropping out
the term with the time derivative of the scalar field $h(t)$ in the effective
action (\ref{eq:DR-model}) and the last two terms in (\ref{eq:W-model}),
i.e. the potential $\tilde{W}$ coinsides with $W$ in this case.

We are looking for the vacua of the model corresponding to the spacetime
$E_{(4+d)} = R^{1} \times S^{3} \times SO(5)/( SU(2) \times U(1))$
with compactified extra dimensions. Physically interesting
solutions are characterized 1) by a large and growing scale $a(t)$ of the
three-dimensional space, and 2) by small and constant or slowly varying scales
$L_{1}$ and $L_{2}$. The latter requirement comes from the fact that
four-dimensional couplings, which are known to be constant to a great
accuracy, are related to multidimensional ones by relations like
(\ref{eq:coupling}), (\ref{eq:kappa}). As it was estimated in ref.
\cite{kolb-perry}, for the yield of primordial $\ ^{4}{\mbox He}$ to
be within acceptable limits the ratio of the value of the size of extra
dimensions $L_{D}$ at the epoch of primordial nucleosynthesis to its present
value $L$ must satisfy the bounds: $0.99 \leq L_{D}/L \leq 1.01$.

Let us fix the scale parameters $L_{10}$ and $L_{20}$ to be such that
\begin{equation}
L_{10}^{2} = \frac{2 v_{0}}{{\cal R}_{0}}, \; \; \; \;
L_{20}^{2} = 2 L_{10}^{2},   \label{eq:L0-relation}
\end{equation}
where $v_{0} = v(\tilde{f};1/2)$, ${\cal R}_{0} = {\cal R}(1/2) =
 5/(16 \pi \kappa )$, $\tilde{f}$
is determined by eq. (\ref{eq:f-minimum}) and the
functions $v(f;\beta)$ and ${\cal R}(\beta)$ are related to the scalar
potential $V(f;L_{1},L_{2})$ (\ref{eq:potential-model}) and the components
$R_{1}$ and $R_{2}$ (\ref{eq:Rp-model}) as follows:
\begin{eqnarray}
\frac{e^{-4 \psi_{1}}}{L_{10}^{4}} v(f;\beta) & = & \frac{1}{8 e^{2}}
 V(f; L_{1},L_{2}),    \label{eq:v-def}  \\
\frac{e^{-2 \psi_{1}}}{L_{10}^{2}} {\cal R} (\beta) & = &
\frac{1}{16 \pi \kappa} ( 4 R_{1}(\beta) \frac{e^{-2\psi_{1}}}{L_{10}^{2}}
 + 2 R_{2}(\beta) \frac{e^{-2 \psi_{2}}}{L_{20}^{2}} ).  \nonumber
\end{eqnarray}
If we now introduce parameter $\Lambda^{(4)}$ defined by
\[ \frac{\Lambda ^{(4)}}{16 \pi \kappa} = \frac{\hat{\Lambda}}{16 \pi \kappa}
 - \frac{{\cal R}_{0}^{2}}{4 v_{0}},      \]
the potential $W(\psi_{1}, \psi_{2}; f)$ in (\ref{eq:W-model}) can be
written as
\begin{equation}
W(\psi_{1},\psi_{2};f) = e^{-6\psi_{1}} 2 \beta \{ \frac{{\cal R}_{0}^{2}}
{4 v_{0}} [ \frac{v(f;\beta)}{v_{0}} e^{-4 \psi_{1}} - \frac{2 {\cal R}(\beta)}
{{\cal R}_{0}} e^{-2\psi_{1}} + 1 ] + \frac{\Lambda ^{(4)}}{16 \pi \kappa} \},
\label{eq:W1}
\end{equation}
where as before $\beta = \exp(2(\psi_{1} - \psi_{2}))/2$.

The evolution of the multidimensional Universe, described by the model, is
determined by the equations of motion for the functions $a(t)$, $\psi_{1}(t)$,
$\psi_{2}(t)$ and $f_{i}(t)$. We will analyze dynamical solutions of these
equations elsewhere. Here we are interested in the vacua of the model, i.e.
extrema of the potential (\ref{eq:W1}). To simplify the problem let us
assume that $f$ takes the value (\ref{eq:f-minimum}) minimizing $v(f;\beta)$
for a given $\beta$, that is $f = \tilde{f}$. Then potential $w(\psi_{1},
\beta) \equiv W(\psi_{1},\psi_{2},\tilde{f})$ takes the form:
\begin{eqnarray}
w(\psi_{1},\beta) & = & e^{-6 \psi_{1}} 2 \beta \{ \frac{{\cal R}_{0}^{2}}
{4 v_{0}} [(1 - e^{-2 \psi_{1}})^{2} + (\beta - \frac{1}{2}) F_{1}(\beta)
e^{-2\psi_{1}} (e^{-2\psi_{1}} - 1)    \nonumber  \\
  & + & (\beta - \frac{1}{2})^{2} F_{2}(\beta) e^{-4\psi_{1}} ] +
\frac{\Lambda ^{(4)}}{16 \pi \kappa} \},    \label{eq:W2}
\end{eqnarray}
\hspace{10cm} for $\beta < 2$;
\begin{eqnarray}
w(\psi_{1},\beta) & = & e^{-6 \psi_{1}} 2 \beta \{ \frac{{\cal R}_{0}^{2}}
{4 v_{0}} [(1 - e^{-2 \psi_{1}})^{2} - (\beta - \frac{1}{2}) F_{1}(\beta)
e^{-2\psi_{1}}    \nonumber  \\
  & + & G_{2}(\beta) e^{-4\psi_{1}} ] +
\frac{\Lambda ^{(4)}}{16 \pi \kappa} \},    \label{eq:W3}
\end{eqnarray}
\hspace{10cm} for $\beta \geq 2$.

\noindent The functions $F_{i}(\beta)$ and $G_{2}(\beta)$ are equal to
\begin{eqnarray}
F_{1}(\beta) & = & \frac{2 \beta + 1}{3 \beta}, \; \; \; \;
F_{2}(\beta) = \frac{2}{3\beta} [ \frac{2}{3} \frac{m-1}{m+1} \beta + 1],
  \nonumber  \\
   G_{2}(\beta) & = & \frac{1}{3} (\frac{3+m}{m+1} + \frac{2m}{m+1}
   \beta^{2}).   \nonumber
\end{eqnarray}
The plot of the potential $w$ as a function of $\psi_{1}$ and
$\psi_{2}$ for $\Lambda ^{(4)} = 0$ and $m=3$ is presented in Fig. 1.

According to eq. (\ref{eq:L-relation}) the line $\beta = 1/2$ on the
$(\psi_{1},\beta)$ - plane corresponds to $SU(4)$-invariant metrics on
the space of extra dimensions realized as
the symmetric space $SU(4)/SU(3) \times U(1)$. We have
\begin{equation}
 w_{symm}(\psi_{1}) \equiv w(\psi_{1}, \frac{1}{2}) = e^{-6\psi_{1}}
[\frac{{\cal R}_{0}^{2}}{4 v_{0}} (1 - e^{-2\psi_{1}})^{2} +
\frac{\Lambda^{(4)}}{16 \pi \kappa} ].   \label{eq:W4}
\end{equation}
A potential of this type was studied in \cite{amendola} - \cite{BKM}.
Let us briefly recall its properties keeping in mind that in our case
$v_{0} > 0$. The shape of $w_{symm}(\psi_{1})$ depends on the value of the
parameter $\Lambda ^{(4)}$. For $ 16 \pi \kappa \Lambda ^{(4)} \geq
5/(12 v_{0})$ the potential does not have any extrema. For
\begin{equation}
 0 \leq 16 \pi \kappa \Lambda ^{(4)} < \frac{5}{12 v_{0}}   \label{eq:Lambda0}
\end{equation}
the potential has a minimum $\psi_{1min}$ and a maximum
$\psi_{1max}$ (see Fig. 2). When the parameters are tuned in such way that
$\Lambda^{(4)}=0$ we get
\begin{eqnarray}
\psi_{1min} & = & 0, \; \; \; \;  w_{symm}(\psi_{1min}) = 0,
                                                   \label{eq:w-min-sym}  \\
\psi_{1max} & = & \frac{1}{2} \ln \frac{5}{3}, \; \; \; \;
   w_{symm}(\psi_{1max}) = \frac{1}{(16 \pi \kappa)^{2}} (\frac{3}{5})^{2}
   \frac{16}{v_{0}}    \label{eq:w-max-sym}
\end{eqnarray}
According to general arguments \cite{coleman} solutions (\ref{eq:w-min-sym})
and (\ref{eq:w-max-sym}) in the class of $SU(4)$-invariant metrics must be also
extrema in the wider class of $SO(5)$-invariant metrics on the space of extra
dimensions realized as the homogeneous space $G/H = SO(5)/SU(2) \times U(1)$.
Indeed, it is easy to check that the potential $w(\psi_{1}, \beta)$
(\ref{eq:W2})-(\ref{eq:W3}) for $\Lambda^{(4)} = 0$ has the minimum
\begin{equation}
(\psi_{1},\beta)_{min} = (0,\frac{1}{2}), \; \; \; \; w(0,\frac{1}{2}) = 0
  \label{eq:w-min}
\end{equation}
and the saddle point
\begin{equation}
 (\psi_{1},\beta)_{saddle} = (\psi_{1max},\frac{1}{2}), \; \; \; \;
 w(\psi_{1max},\frac{1}{2}) = \frac{1}{(16 \pi \kappa)^{2}} (\frac{3}{5})^{2}
   \frac{16}{v_{0}}    \label{eq:w-saddle}
\end{equation}
with $\psi_{1max}$ given by eq. (\ref{eq:w-max-sym}).  One can also show by
straightforward calculations that there are no other extrema of $w(\psi_{1},
\beta)$ corresponding to cosmological solutions $E_{(10)} = R^{1} \times S^{3}
\times SO(5)/SU(2) \times U(1)$ with compactified spacelike extra
dimensions. We would like to notice here that spontaneous compactification
solutions for $E_{(10)} = M^{4} \times SO(2k+1)/SU(k) \times U(1)$, $(k \geq
2)$ with both spacelike and timelike extra dimensions were found and studied in
\cite{KMV}. In that case there is only one solution with spacelike extra
dimensions, it corresponds to (\ref{eq:w-min}) in the limit of infinite
size $a = \infty$ of three dimensions. The relation between compactifying
cosmological solutions, corresponding to four-dimensional spacetime
$M_{(4)}$ being of the closed Friedman-Robertson-Walker type, and spontaneous
compactification solutions, corresponding to the four-dimensional spacetime
being the Minkowski spacetime, is discussed in the Appendix.

We see now that the parameters $L_{10}$ and $L_{20}$ given by eq.
(\ref{eq:L0-relation}) are characteristic scales of the compactified
internal space corresponding to the minimum (\ref{eq:w-min}) of the
potential when $\Lambda^{(4)} = 0$.

Let us discuss now the potential $w(\psi_{1},\beta) \equiv
w(\psi_{1},\psi_{2};\tilde{f})$ for small $\Lambda^{(4)}$, namely for
$| 16 \pi \kappa \Lambda^{(4)}| \ll 1$ (see below). It is clear that
its shape deforms only slightly in comparison with the case
$\Lambda ^{(4)} = 0 $. It turns out that its minimum and its maximum are
located on the line $\beta = 1/2$ for all values of $\Lambda ^{(4)}$
satisfying condition (\ref{eq:Lambda0}). Thus, the potential
reaches its minimum at
\[ \left( \psi _{1 min} \approx 16 \pi \kappa \Lambda ^{(4)} \frac{3
          v_{0}}{25}, \beta _{min} = \frac{1}{2} \right)  \]
and
\[ w(\psi _{1min}, \beta _{min}) = \frac{\Lambda ^{(4)}}{16 \pi \kappa}
 \left( 1 + O(16 \pi \kappa \Lambda ^{(4)}) \right).  \]
We see that parameter $\Lambda ^{(4)}$ is the cosmological constant of
the reduced theory in four dimensions.

There are two directions in the $(\psi_{1},\psi_{2})$-plane along
which the potential approaches constant values when
$|\psi_{1}|$, $|\psi_{2}| \rightarrow \infty$:
\begin{eqnarray*}
a) & \psi_{1} = 0, & \; \; \; \psi_{2} \rightarrow +\infty
   \; \; \; \; \mbox{or}   \; \; \; \;             \beta \rightarrow 0   \\
   &  & w(0,\beta)    \sim \frac{1}{(16 \pi \kappa)^{2}} \frac{12}{v_{0}} \\
b) & \psi_{1} \rightarrow \infty, & \; \; \; \psi_{2}=-\frac{2}{3}\psi_{1}
 \; \; \; \;  \mbox{or} \; \; \; \; \beta = \frac{1}{2} e^{10 \psi_{1} /3},\\
   &  & w(\psi_{1}, \frac{1}{2} e^{10 \psi_{1}/3}) \sim
        \frac{1}{(16 \pi \kappa)^{2}} \frac{6m}{m+1} \frac{1}{v_{0}}
\end{eqnarray*}

If we introduce variables $\Psi = \sqrt{\psi_{1}^{2} + \psi_{2}^{2}}$ and
$\alpha = \arctan (\psi_{2} / \psi_{1})$, then when $\Psi \rightarrow \infty$
the potential grows exponentially in the sector $(-\arctan \frac{2}{3}) <
\alpha < \frac{\pi}{2}$ and decreases exponentially to zero in the sector
$\frac{\pi}{2} < \alpha < 2\pi - \arctan \frac{2}{3}$. There is
a hollow centered at the minimum of the potential; if $\Lambda ^{(4)} = 0$ it
reaches zero at $\psi_{1} = \psi_{2} = 0$ (see (\ref{eq:w-min})). The hollow
is separated from the region of exponential decreasing of the potential
by a ridge joining the rays $\alpha = (\pi/2)$ and $\alpha = -\arctan (2/3)$
with the pass at the saddle point (\ref{eq:w-saddle}).

Thus, we see that for zero temperature the compactifying solution exists which
is classically stable. The interesting feature is that the minimum of the
potential of the model has higher symmetry
than a general configuration, namely it corresponds to the $SU(4)$-invariant
metric satisfying (\ref{eq:L-relation}) (or $\beta = 1/2$). We will
show in the next section this property of the vacuum is not affected
by non-zero temperature contributions.

When $\Lambda^{(4)} > 0$ the minimum is semiclassically unstable because of
non-zero probability of quantum tunneling to the region of exponential
vanishing of the potential. This region corresponds to unlimited growth of the
scales of the internal space and thus to decompactification of extra
dimensions. However, since the multidimensional cosmological constant must
be fine-tuned in the appropriate way so that the four-dimensional cosmological
constant satisfies the bound $16 \pi \kappa |\Lambda ^{(4)}| < 10^{-120}$,
the tunelling rates of the dilaton fields $\psi_{1}$ and $\psi_{2}$ appear
to be extremely small and the lifetime of the metastable vacuum near
$\psi_{1} = \psi_{2}=0$ exceeds the age of the Universe \cite{BKM}.

In the next section we will study the potential of the dilaton fields
$\psi_{1}$ and $\psi_{2}$ when temperature effects in the radiation dominated
period, which follows inflation, are taken into account.

\section{Stability of compactification after inflation}

The analysis carried out in \cite{KRT} and the arguments presented in
\cite{BKM} show that in the case of a symmetric internal space
the expansion of three spatial dimensions provided by the dilaton field
is only of the power type $a(t) \sim t^{p}$ with $p<1$ and that there
is no any natural mechanism of achieving inflation driven by the dilaton.
Similar considerations are also true in our case, and therefore in order to
achieve a successful period of inflation one needs necessarily to add an
inflaton sector. Thus we assume that the inflation of the macroscopic
dimensions was driven by some additional inflaton field and the process of
shrinking of extra dimensions to a small size took place at the period
that followed the accelerated
expansion of the spatial dimensions of $M_{(4)}$ and subsequent reheating. In
this section we consider the process of spontaneous compactification of
extra dimensions during this period when the Universe was radiation dominated.
Similarly to the paper \cite{BKM} we describe this situation by assuming
that the main contribution to the temperature comes from nonvanishing
external space components of the gauge field $A_{\mu}$ $(\mu = 0,1,2,3)$.
As it has been already explained in Sect. 3 in accordance with the type of
the cosmological model considered here, we assume also that all physical
observables are spatially homogeneous. This means
that the fields $\psi _{i}$ and $f_{i}$ depend on $t$ only and $A_{\mu}$ is
symmetric under the action of $SO(4)$. For the sake of simlicity
we will restrict ourselves to the case $m=3$, i.e. the gauge group of the
initial $10$-dimensional theory is $K=SU(5)$ and the gauge group after
dimensional reduction to four dimensions is $C=SU(3) \times U(1)$. The action
of the effective theory in the cosmological setting was derived in Sect. 3
and is given by eqns. (\ref{eq:DR-model}) - (\ref{eq:W-model}).
The contribution of the four dimensional
components of the gauge field to the energy density is given by
\begin{equation}
\rho (\psi_{p},f,h) = \frac{1}{a^{4}} \frac{3}{4e^{2}} e^{4\psi_{1}+2\psi_{2}}
 [a^{2} \dot{h}^{2} + (1-h^{2})^{2}] + \frac{2}{a^{2}} L_{10}^{2}
 e^{-2\psi_{1}} (1+h)^{2} |f|^{2}.  \label{eq:3-contribution}
\end{equation}
After inflation the vacuum energy density is transformed into
thermal energy $\rho$ via large nonstatic vacuum expectation values for $h$.
As before we assume that the internal components of the gauge field,
described by $f$, are given by the static minimum $\tilde{f}$
(\ref{eq:f-minimum}). It was shown in \cite{BKM} that this assumption is
self-consistent. Then the dynamics of the dilaton fields is governed by the
potential
\[ \tilde{W} (\psi_{p},f_{min},h) = W(\psi_{p}, f_{min}) + \rho(\psi_{p},
    f_{min},h).  \]
During inflation the scale factor $a(t)$ grows exponentially. Therefore the
last term in (\ref{eq:3-contribution}) does not change the minimum $\tilde{f}$
essentially. By introducing the temperature associated with the gauge field
$A_{\mu}(x)$, $T \sim 1/a$, we obtain
\[ \rho(\psi_{p},f,h) = \sigma T^{4} e^{4\psi_{1} + 2\psi_{2}}  \]
with $\sigma$ being a constant of the order one. The potential describing the
dymanics of $\psi_{1}$ and $\psi_{2}$ now is equal to
\begin{equation}
\tilde{w} (\psi_{1},\beta) = w (\psi_{1},\beta) + \frac{\sigma T^{4}}
  {2\beta} e^{6\psi_{1}},     \label{eq:wT}
 \end{equation}
where $w(\psi_{1},\beta)$ is given by (\ref{eq:W2}), (\ref{eq:W3}).
For non-zero temperature it grows exponentially
when $\Psi \equiv \sqrt{\psi^{2}_{1} + \psi^{2}_{2}}$ goes to infinity,
for $\Lambda ^{(4)} = 0$ and $m=3$ its shape is shown in Fig. 3.
When $T \neq 0$ the potential $\tilde{w}$ has more extrema than in the
case $T=0$ and we are
going to discuss them now. For the analysis that follows it is convenient to
introduce the notation
\[  {\cal A} = \frac{{\cal R}_{0}^{2}}{4 v_{0}}.    \]

After the inflation the temperature satisfies $T \leq 10^{-5}/ \sqrt{16 \pi
\kappa}$. Therefore, in the region of finite $\Psi$ the second term in eq.
(\ref{eq:wT}) leads to small shifts of the positions of the extrema of
$w$ only and the types of the extrema remain to be the same. Thus,
the potential $\tilde{w}$ has a saddle point $(\psi_{1},
\beta)_{saddle}$, with $\beta = 1/2$ and $\psi_{1}$ differring from
$(\psi_{1})_{max}$ in (\ref{eq:w-saddle}) by small temperature corrections,
and a minimum with
\begin{eqnarray}
 \psi_{1min} & = & -\frac{\sigma T^{4}}{4 {\cal A}}, \; \; \;
\beta_{min}= \frac{1}{2},   \label{eq:wT-min}  \\
 \tilde{w}_{min} & \equiv & \tilde{w}(\psi_{1min},\beta_{min})
 \simeq \sigma T^{4}              \nonumber
\end{eqnarray}
(cf. (\ref{eq:w-min})). Notice that even for non-zero temperature the saddle
point and the minimum are located on the line $\beta = 1/2$ corresponding to
invariant metrics with higher symmetry, namely to $SU(4)$-invariant metrics
of the symmetric space $SU(4)/SU(3) \times U(1)$. Another minimum,
which exists when $T \neq 0$, is located also on this line but far from
the origin of the $(\psi_{1},\psi_{2})$-plane:
\begin{eqnarray}
\psi_{1a} & \simeq & \frac{1}{12} \ln \frac{3 {\cal A}}{\sigma T^{4}},
\; \; \; \beta_{a} = \frac{1}{2},  \label{eq:wT-min-a}   \\
\tilde{w}_{a} & \equiv & \tilde{w}(\psi_{1a},\beta_{a}) \simeq 2 T^{2}
\sqrt{\sigma {\cal A}}(1 - (\frac{\sigma T^{4}}{{\cal A}})^{1/6}).   \nonumber
\end{eqnarray}
The minima analogous to (\ref{eq:wT-min}) and (\ref{eq:wT-min-a}) were found
and analyzed in the paper \cite{BKM} where the case of symmetric space of
extra dimensions was considered.

The potential $\tilde{w}$ has two more extrema which are located away from
the line $\beta = 1/2$ of $SU(4)$-invariant metrics. One of them is a
saddle point
\begin{eqnarray}
\psi_{1b} & \simeq & \frac{1}{12} \ln \frac{4 {\cal A}}{\sigma T^{4}},
\; \; \; \beta _{b} \simeq 1 + (\frac{\sigma T^{4}}{4 {\cal A}}),
                                                      \label{eq:wT-sad-b}  \\
\tilde{w}_{b} & \equiv & \tilde{w}(\psi_{1b},\beta_{b}) \simeq
2 T^{2} \sqrt{\sigma {\cal A}} ( 1 - \frac{5}{4} (\frac{\sigma T^{4}}
{4 {\cal A}})^{1/6}),    \nonumber
\end{eqnarray}
and another is a minimum
\begin{eqnarray}
\psi_{1c} & \simeq & \frac{1}{8} \ln (\frac{112}{81} \frac{{\cal A}}{\sigma
T^{4}}), \; \; \; \beta _{c} \simeq \frac{4}{9} (\frac{7 {\cal A}}
{\sigma T^{4}})^{1/4},                              \label{eq:wT-min-c}  \\
\tilde{w}_{c} & \equiv & \tilde{w}(\psi_{1c},\beta_{c}) \simeq
\frac{2}{3} T^{2} \sqrt{7 {\cal A} \sigma}.   \nonumber
\end{eqnarray}
Notice that the saddle point (\ref{eq:wT-sad-b}) exceeds the minimum
(\ref{eq:wT-min-a}) in the subleading order in $T$: $\tilde{w}_{b} -
\tilde{w}_{a} \sim T^{8/3}$. Another feature of this saddle point of the
potential is that $\beta_{b} \rightarrow 1$ when $T \rightarrow 0$ and the
invariant metric on the space of extra dimensions asymptotically
approaches the second Einstein metric with $L_{1} = L_{2}$ (see Sect. 3).

We see that $\tilde{w}_{min}$ is the absolute minimum of the potential. So,
the configuration (\ref{eq:wT-min}) is the true vacuum with compactified
extra dimensions and it is semiclassically stable when the temperature is
non-zero.

The second minimum (\ref{eq:wT-min-a}) corresponds to the metastable vacuum
with relatively large sizes $L_{1}$ and $L_{2}$ of extra dimensions, which
grow to infinity when $T \rightarrow 0$. This means decompactification of
extra dimensions. If the Universe appeared to be
in this state after inflation, it would tend to tunnel to the vacuum
$(\psi_{1min}, 1/2)$ with compactified extra dimensions. However, as it was
estimated in \cite{BKM} for the symmetric case, the lifetime of the
metastable vacuum $(\psi_{1a},\beta_{a}=1/2)$ exceeds the present age of
the Universe and thus the situation when the system arrives to this vacuum
as result of the dynamical evolution is unsatisfactory from the
phenomenological point of view.

The minimum $(\psi_{1c},\beta_{c})$ is determined by the part of the potential
described by the expression (\ref{eq:W3}). It has rather interesting
feature. As it can be easily seen from  (\ref{eq:wT-min-c}) when the
temperature vanishes the scale $L_{1}$ increases to infinity whereas the
scale $L_{2}$ approaches a constant value: $L_{2} \rightarrow \sqrt{3}
L_{20} /2$, where $L_{20}$ is given by (\ref{eq:L0-relation}). This means that
if the system rests in this metastable vacuum (similar to the case of the
minimum $(\psi_{1a},1/2)$ its lifetime exceeds the lifetime of the Universe)
four out of six extra dimensions, corresponding to the subspace ${\cal M}_{1}$
in the decomposition (\ref{eq:decomp-M}), decompactify whereas the
two-dimensional subspace, corresponding to ${\cal M}_{2}$, remains to be
compact. Thus, in the limit of small temperature this vacuum effectively
describes the Universe with eight macroscopic dimensions whereas the true
vacuum (\ref{eq:wT-min}) corresponds to the one with four macroscopic
dimensions. Depending on the initial conditions the system ends up its
evolution in one of the minima found above and, therefore, depending on the
initial conditions that or another number of macroscopic dimensions
develops dynamically in the Universe. This mechanism might be important
for describing topological transitions in the space of macroscopic
dimensions in the early Universe.

\section{Conclusions}

In the present paper we have considered the problem of compactification after
inflation, i.e. after reheating in the Einstein-Yang-Mills
theory with extra dimensions being the non-symmetric homogeneous space
$G/H = SO(5)/SU(2) \times U(1)$. Our aim was to analyze the model
in the cosmological setting, so we have imposed the condition that the
relevant fields were spatially homogeneous. We have taken into account
non-vanishing components of the gauge field in four dimensions and this
allowed us to introduce the temperature into our analysis.

The space of all $SO(5)$-invariant metrics on the space of extra
dimensions in the model is characterized by two scales, $L_{1}$
and $L_{2}$. In this sence our results are complementary to those of
ref.\cite{BKM} where the case of symmetric space of extra dimensions was
considered. We studied configurations corresponding
to compactifying solutions of the Einstein-Yang-Mills system.
For non-vanishing temperature there is a static configuration with four
macroscopic dimensions represented by the Friedman-Robertson-Walker spacetime
$R^{1} \times S^{3}$ and six compact dimensions. We found that this
solution is the absolute minimum of the effective potential.
It describes a metric with higher symmetry, namely the $SU(4)$-metric with
$L_{2}^{2} = 2 L_{1}^{2}$, and
the geometry of the space of extra dimensions equal to that of the
symmetric space $SU(4)/SU(3) \times U(1)$.

Furthermore, we find that at a non-zero temperature there are two other
minima of the effective potential which are semiclassically unstable.
One minimum corresponds to the symmetric space of extra dimensions with
all six additional dimensions being decompactified when the temperature
goes to zero. Another minimum corresponds to the non-symmetric space of
extra dimensions and when $T \rightarrow 0$ four of additional
dimensions decompactify whereas two of them remain compact.
The rates of transitions from these vacua to the true compactified vacuum
are very low and, thus, if after the inflation the Universe had been trapped
in one of them  this would be in contradiction with the standard cosmological
scenario. However, this model gives an example when the number of macroscopic
and compact dimensions and their topology depend on the type the vacuum in
which the Universe was trapped after the inflation. This suggests a type
of the mechanism which might be responsible for topological transitions
in the early Universe. In our model the vacuum
with four macroscopic dimensions and six-dimensional compact symmetric
homogeneous space of extra dimensions is the true vacuum of the system.

\section {\bf Acknowledgements}

It is a pleasure for us to thank J.M. Mour\~{a}o and E. Verdaguer
for illuminative and critical discussions of the problems addressed in
this paper. Yu. K. acknowledges support from Direcci\'{o}n General de
Investigaci\'{o}n Cient{\ii}fica y T\'{e}cnica (sabbatical grant
SAB 92 0267) during part of this work and thanks the Department
d'ECM de la Universidad de Barcelona for its warm hospitality.

\newpage

\section*{Appendix}

In this Appendix we discuss the relation between the spontaneous
compactification solutions with spacetime $M^{4} \times G/H$, where
$M^{4}$ is the Minkowski spacetime, and compactifying solutions in a
cosmological setting with spacetime $R^{1} \times S^{3} \times G/H$
for Einstein-Yang-Mills theories \footnote{The authors benefited very much
from pleasant and illuminative discussions of this problem with
J. Mour\~{a}o}. For the purpose of simplicity we will restrict
ourselves to the case of symmetric space $G/H$, the generalization to
the non-symmetic case is straightforward. We also assume that
four-dimensional components of the gauge field are equal to zero.

For $(4+d)$-dimensional spacetime $M_{(4)} \times G/H$ with arbitrary
four-dimensional $M_{(4)}$ the Einstein equations read
\begin{eqnarray}
\hat{R}_{\mu \nu} - \frac{1}{2} g_{\mu \nu} \hat{R} & = &
 - \alpha g_{\mu \nu} Tr \hat{F}^{2} + \hat{\Lambda} g_{\mu \nu} \nonumber \\
\hat{R}_{mn}      - \frac{1}{2} g_{mn}  \hat{R}     & = &
 - \alpha (g_{mn} Tr \hat{F}^{2} - 4 Tr \hat{F}_{mk} \hat{F}^{k}_{n}) +
 \hat{\Lambda} g_{mn},              \nonumber
\end{eqnarray}
where $\alpha = 4 \pi \kappa / e^{2}$ and $\mu, \nu = 0,1,2,3$, $m,n = 4,
\ldots , 3+d$.  For invariant metric and symmetric gauge fields on $G/H$ this
system can be easily transformed into the pair of compactification equations:
\begin{eqnarray}
  R^{(4)} & = & - \frac{2}{d+2} \alpha Tr \hat{F}^{2} -
  \frac{8 \hat{\Lambda}}{d+2}                \label{eq:sc-1}   \\
  R^{(d)} & = & \frac{2(d+4)}{d+2} \alpha Tr \hat{F}^{2} -
       \frac{2d}{d+2} \hat{\Lambda}         \label{eq:sc-2}
\end{eqnarray}
where $R^{(4)}$ and $R^{(d)}$ are scalar curvatures of four-dimensional
spacetime and the space of extra dimensions respectively. Obviously,
$\hat{R} = R^{(4)} + R^{(d)}$. Now we take into account that for $G$-invariant
metric and $G$-symmetric gauge field on the symmetric space $G/H$
\[  R^{(d)} = \frac{k}{L^{2}}, \; \; \;	\; Tr \hat{F}^{2} =
    \frac{v(f)}{L^{4}},   \]
where $L$ is the scale of the space of extra dimensions, $k$
is some constant and $v(f)$ is the potential of the scalar field $f$ of the
reduced theory (cf. (\ref{eq:potential-gen}), (\ref{eq:v-def})). From
the Yang-Mills equation it follows that the scalar field must be equal to
one of the extrema of the potential \cite{dimred2}, \cite{KMRV-review} -
\cite{KMV}, we choose it to be equal to the minimum of the potential,
$f = \tilde{f}$, as in Sect. 4. The important point here is whether
the potential $v(f)$ equals zero or not at the minimum. This property
of the potential depends on the homomorphism $\tau$ from the isotropy group
$H$ into the gauge group $K$ of the multidimensional theory (see Sect. 2).
We will assume here that $\tilde{v} = v(\tilde{f}) \neq 0$, otherwise the
Einstein-Yang-Mills model does not have spontaneous compactification
solutions.

Let us consider first the spontaneous compactification solution
when four-dimensional spacetime is the Minkowski space $M^{4}$. Then
$R^{(4)} = 0$ and the first equation (\ref{eq:sc-1}) gives a relation
between the multidimensional cosmological constant $\hat{\Lambda}$
and $\tilde{v}$.
This relation is exactly the condition of the vanishing of the
four-dimensional cosmological constant in the reduced theory. The second
equation (\ref{eq:sc-2}) then gives the value of the radius of
compactification:
\begin{equation}
     L_{SC}^{2} = \frac{4 \alpha \tilde{v}}{k}.        \label{eq:sc-solution}
\end{equation}

Now let us study similar problem but when four-dimensional spacetime is the
closed Friedman-Robertson-Walker Universe $M_{(4)} = R^{1} \times S^{3}$ with
the metric (\ref{eq:4-metric}). In this case we look for solutions with
positive and (in general) non-constant radius $a(t)$ of three-dimensional
space and constant size $L$ of the compact space of extra dimensions and
consider $\hat{\Lambda}$ as a parameter. The second equation
(\ref{eq:sc-2}) gives us two solutions $L_{-}$ and $L_{+}$ which depend on
$\hat{\Lambda}$. It is easy to
check that, if $0 < (- \hat{\Lambda}) < k^{2}/(-16 \alpha \tilde{v} d(d+4)
(d+2)^{2})$, one of them is a minimum (we denote it $L_{-}$) and another is a
maximum of the effective potential
\[ W(L) = (\frac{L}{L_{0}})^{-d} (-\frac{k}{L^{2}} \frac{2 \alpha \tilde{v}}
 {L^{4}} - 2 \hat{\Lambda} )        \]
of the reduced theory analogous to (\ref{eq:potential-W}). If we impose the
additional condition $W_{-} \equiv W(L_{-}) = 0$, which means vanishing of
the four-dimensional cosmological constant in the vacuum $L_{-}$, then
we will find that the multidimensional cosmological constants must be fine
tuned to be $\hat{\Lambda} = k^{2}/(16 \alpha \tilde{v})$ and
\begin{eqnarray}
  L_{-}^{2} & = &  \frac{4 \alpha \tilde{v}}{k},    \label{eq:s1-solution} \\
  L_{+}^{2} & = &  \frac{4 \alpha \tilde{v} (d+4)}{kd}.
  \label{eq:s2-solution}
\end{eqnarray}
This result was obtained in a number of papers, see e.g. \cite{KRT},
\cite{BKM}. Notice that the value of the radius of extra dimensions in the
compactification vacuum given by eq. (\ref{eq:s1-solution}) coincides with
the spontaneus compactification solution (\ref{eq:sc-solution}). Then,
from the first equation (\ref{eq:sc-1}) it follows that for $L = L_{-}$
the scalar curvature of four-dimensional spacetime $R^{(4)}=0$. This
means that either $M_{(4)}$ is the Minkowski space (this case was discussed
above) or $\dot{a} \neq 0$. From (\ref{eq:sc-1})
it also follows that the solution $L_{+}$ exists for $R^{(4)} > 0$ only.

We see that if the condition of the vanishing of the four-dimensional
cosmological constant is imposed, the spontaneous compactification solution
for $L$ always survives in a cosmological setting. But in the latter case we
have additional solution with positive scalar curvature of four-dimensional
spacetime, which does not exist when $M_{(4)} = M^{4}$.

\newpage

\section*{Figure captions}

\begin{description}
  \item[Fig. 1] Potential $w(\psi_{1},\psi_{2})$ of the
                model obtained by dimensional reduction of the
		ten-dimensional Einstein-Yang-Mills theory with the space
		of extra dimensions $SO(5)/SU(2) \times U(1)$. Zero
		temperature case. The plotted function is given
		by eqns. (\ref{eq:W2}), (\ref{eq:W3}) (recall that
		$\beta = e^{2(\psi_{1} - \psi_{2})} / 2$) with
		$\Lambda ^{(4)} = 0$ and $m=3$.
  \item[Fig. 2] Function $w_{symm}(\psi_{1})$ given by eq.
                (\ref{eq:W4}) for $\Lambda^{(4)} = 0$. This
		plot is the section of the surface in Fig. 1 for $\beta = 1/2$.
  \item[Fig. 3] Potential $\tilde{w}(\psi_{1},\psi_{2})$ of the model for
                non-zero temperature (see eq. (\ref{eq:wT})) for
		$\Lambda ^{(4)} = 0$, $m=3$, $\sigma v_{0} = 1$ and
		$T = 10 ^{-3} / \sqrt{16 \pi \kappa}$.

\end{description}


\begin{thebibliography}{99}
\bibitem[1]{KK} Th. Kaluza, {\em Sitzungsber. Preuss. Akad. Wiss. Math.
Phys.} {\bf K1} (1921) 966. \\
O. Klein, {\em Z. Phys.} {\bf 37} (1926) 895.

\bibitem[2]{kerner} J. Rayski, {\em Acta Phys. Polon.} {\bf 27} (1965)
947; \\
R. Kerner, {\em Ann. Inst. H. Poinc.} {\bf A9} (1968) 143; \\
A. Trautman, {\em Rep. Math. Phys.} {\bf 1} (1970) 29; \\
Y.M. Cho, {\em J. Math. Phys.} {\bf 16} (1975) 2029.

\bibitem[3]{salam} A. Salam and J. Strathdee, {\em Ann. Phys.} {\bf 141}
(1982) 316.

\bibitem[4]{DNP} M.J. Duff, B.E.W. Nilsson and C.N. Pope, {\em Phys. Rep.}
{\bf 130} (1981) 1.

\bibitem[5]{coquer} R. Coquereaux and A. Jadczyk, {\em Riemannian Geometry,
Fiber Bundles, Kaluza-Klein Theories and All That ...} Lecture Notes in
Physics, Vol. 16 (World Scientific, Singapore, 1988).

\bibitem[6]{sc-quantum} S.R. Huggins, G. Kunstatter, H.P. Leivo and D.J. Toms,
{\em Nucl. Phys.} {\bf B301} (1988) 627; \\
I.L: Buchbinder and S.D. Odintsov, {\em Fortschr. der Phys.} {\bf 37} (1989)
225;\\
H.T. Cho and R. Kantowski, {\em Phys. Rev. Lett.} {\bf 67} (1991) 422.

\bibitem[7]{witten} E. Witten, in Shelter Island II: Proceedings of the
1983 Shelter Island Conference on Quantum Field Theory and the Fundamental
Problems of Physics, eds. R. Jackiw et al. (MIT Press, Cambridge, Mass.,
1985).

\bibitem[8]{horv-pal} Z. Horvath, L. Palla, E. Cremmer and J. Scherk,
{\em Nucl. Phys.} {\bf B127} (1977) 57; \\
N.S. Manton, {\em Nucl. Phys.} {\bf B158} (1979) 141; \\
P. Forgacs and N.S. Manton, {\em Commun. Math. Phys.} {\bf 72} (1980) 15.

\bibitem[9]{dimred1} G. Chapline and R. Slansky, {\em Nucl. Phys.}
{\bf B209} (1982) 461; \\
I.P. Volobujev and G. Rudolph, {\em Theor. Math. Phys.} {\bf 62}
(1985) 261; \\
K. Pilch and A.N. Schellekens, {\em Nucl. Phys.} {B256} (1985) 109; \\
I.P. Volobujev and Yu.A. Kubyshin, {\em Theor. Math. Phys.} {\bf 68} (1986)
788, 855; \\
F.A. Bais, K.J. Barnes, P. Forgacs and G. Zoupanos, {\em Nucl. Phys.}
{\bf B263} (1986) 557; \\
N.G. Kozimirov and I.I. Tkachev, {\em Z. Phys.} {\bf C 36} (1987) 83; \\
 K. Farakos, G. Koutsoumbas, M. Surridge and G. Zoupanos,
{\em Nucl. Phys.} {\bf B291} (1987) 128; \\
Yu.A. Kubyshin, J.M. Mour\~{a}o and I.P. Volobujev, {\em Int. J. Mod. Phys.}
{\bf A4} (1989) 151.

\bibitem[10]{dimred2} M. Surridge, {\em Z. Phys.} {\bf C 37} (1987) 77; \\
Yu.A. Kubyshin, J.M. Mour\~{a}o and I.P. Volobujev, {\em Phys. Lett.}
{\bf B203} (1988) 349; \\
A. Nakamura and K. Shiraishi, Report No.INS-Rep. 795 (1989).

\bibitem[11]{FKSZ}  K. Farakos, G. Koutsoumbas, M. Surridge and G. Zoupanos,
{\em Phys. Lett.} {\bf B191} (1987) 135.

\bibitem[12]{KMV} Yu.A. Kubyshin, J. Mour\~{a}o and I.P. Volobujev,
{\em Nucl. Phys.} {\bf B322} (1989) 531.

\bibitem[13]{KMRV-review} Yu.A. Kubyshin, J.M. Mour\~{a}o, G. Rudolph and
I.P: Volobujev, {\em Dimensional Reduction of Gauge Theories, Spontaneous
Compactification and Model Building}, Lecture Notes in Physics, Vol. 349
(Springer-Verlag, Berlin, 1989).

\bibitem[14]{KZ-review} D. Kapetanakis and G. Zoupanos, {\em Phys. Rep.}
{\bf C219} (1992) 1.

\bibitem[15]{GSW} M.B. Green, J.N. Schwarz and E. Witten, {\em Superstring
Theory}. Vol. I, II. (Cambridge University Press, Cambridge, 1987).

\bibitem[16]{cremmer} E. Cremmer and J. Scherk, {\em Nucl. Phys.} {\bf B118}
(1977) 61. \\
J.F. Luciani, {\em ibid} {\bf B135} (1978) 11.

\bibitem[17]{dynsc} P.G.O. Freund, {\em Nucl. Phys.} {\bf B209} (1982)
146; \\  V.A. Rubakov and M.E. Shaposhnikov, {\em Phys. Lett.} {\bf 125B}
(1983) 139; \\ Q. Shafi and C. Wetterich, {\em ibid.} {\bf 129B} (1983)
387; \\ E.W. Kolb and R. Slansky, {\em ibid.} {\bf 135B} (1984) 378; \\
A.B. Henriques, {\em Nucl. Phys.} {\bf B277} (1986) 621; \\
A.B. Henriques, A.R. Liddle and R.G. Moorhouse, {\em ibid.} {\bf B311}
(1989) 719.

\bibitem[18]{kolb-perry} E.W. Kolb, M.J. Perry and T.P. Walker,
{\em Phys. Rev. } {\bf D33} (1986) 869.

\bibitem[19]{maeda} K. Maeda, {\em Phys. Lett.} {\bf 186B} (1987) 33.

\bibitem[20]{amendola} L. Amendola, E.W. Kolb, M. Litterio and
F. Occhionero, {\em Phys. Lett.} {\bf D42} (1990) 1944.

\bibitem[21]{KRT} Yu.A. Kubyshin, V.A. Rubakov and I.I. Tkachev,
{\em Int. J. Mod. Phys.} {\bf A 4} (1989) 1409.

\bibitem[22]{BKM} O. Bertolami, Yu.A. Kubyshin and J.M. Mour\~{a}o,
{\em Phys. Rev.} {\bf D45} (1992) 3405.

\bibitem[23]{stability} S. Randjbar-Daemi, A. Salam and J. Strathdee, {\em
Nucl. Phys.} {\bf B214} (1983) 491; {\em Phys. Lett.} {\bf 124B} (1983)
345; \\
A.N. Schellekens, {\em Nucl. Phys.} {\bf B248} (1984) 704; \\
P. Forgacs, Z. Horvath and L. Palla, {\em Phys. Lett.} {\bf 147B} (1984) 311.

\bibitem[24]{KN} S. Kobayashi and K. Nomizu, {\em Foundations of Differential
Geometry.} Vols. I and II (Interscience Publ., NY, 1969).

\bibitem[25]{BMPV} O.Bertolami, J.M. Mour\~{a}o, R.F. Picken and
I.P. Volobujev, {\em Int. J. Mod. Phys.} {\bf A 6} (1991) 4149. \\
P.V. Moniz and J.M. Mour\~{a}o, {\em Class. Quant. Grav.} {\bf 8} (1991) 1815.

\bibitem[26]{jadczyk} A. Jadczyk. Wroclaw University preprint ITP-UWr 84/615
(1984).

\bibitem[27]{mourao} J.M. Mour\~{a}o, JINR preprint E2-88-155 (Dubna, JINR,
1988).

\bibitem[28]{ziller} W.Ziller, {\em Math. Ann.} {\bf 259} (1982) 351.

\bibitem[29]{coleman}S.Coleman, in {\em Proc. of Int. School of Subnuclear
Physics 1975, Part A}, ed. A. Zichichi (Plenum Press, London, 1977). \\
M.J. Duff and C.N. Pope, {\em Nucl. Phys.} {\bf B255} (1985) 355.

\end{thebibliography}
\end{document}